# Noncontiguous I/O through PVFS

Avery Ching, Alok Choudhary, Wei-keng Liao
*Center for Parallel and Distributed Computing*
*Northwestern University*
*Evanston, IL 60208*
{aching, choudhar, wkliao}@ece.northwestern.edu

Rob Ross and William Gropp
*Mathematics and Computer Science Division*
*Argonne National Laboratory*
*Argonne, IL, 60439*
{rross, gropp}@mcs.anl.gov

**Abstract**

*With the tremendous advances in processor and memory technology, I/O has risen to become the bottleneck in high-performance computing for many applications. The development of parallel file systems has helped to ease the performance gap, but I/O still remains an area needing significant performance improvement. Research has found that noncontiguous I/O access patterns in scientific applications combined with current file system methods to perform these accesses lead to unacceptable performance for large data sets. To enhance performance of noncontiguous I/O, we have created list I/O, a native version of noncontiguous I/O. We have used the Parallel Virtual File System (PVFS) to implement our ideas. Our research and experimentation shows that list I/O outperforms current noncontiguous I/O access methods in most I/O situations and can substantially enhance the performance of real-world scientific applications.*

## 1. Introduction

The low cost and scalability of cluster computing have made it the most popular platform today. Nevertheless, as on traditionally massively parallel computers, I/O remains a challenge. The Parallel Virtual File System (PVFS), a high-performance parallel file system for Linux clusters, provides a starting point for I/O solutions in this environment [2].

Scientific computing often requires noncontiguous access of small regions of data [1][4][7][11][12]. Traditionally, parallel file systems perform multiple contiguous I/O operations to satisfy these types of requests, resulting in a large I/O request processing overhead.

This paper describes a method for high-performance noncontiguous data access through our implementation called *list I/O*. We chose to implement list I/O using PVFS. This paper presents results from three different benchmarks: an artificial benchmark, an I/O simulation of the FLASH code benchmark, and an I/O simulation of a tiled visualization application. Our results show that in most cases, list I/O outperforms traditional noncontiguous methods by up to two orders of magnitude. We first give an overview of PVFS in Section 2. In section 3 we then offer an in depth analysis of the noncontiguous I/O problem. Section 4 describes the machine configuration we used and our experimental results. Section 5 summarizes our work and briefly outlines future research.

## 2. PVFS Overview

To develop optimizations for noncontiguous access, we used the Parallel Virtual File System to implement our ideas [2]. PVFS is the leading parallel file system for Linux cluster computing and has enabled low-cost clusters of high-performance PCs to address parallel applications with large-scale I/O needs [6]. An example PVFS system configuration is shown in Figure 1.

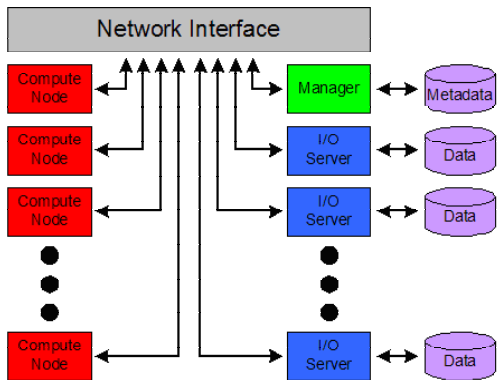

Figure 1. **Example PVFS setup**. *In PVFS, metadata is stored on the manager node. File data is striped across a user-specified number of I/O servers. Compute nodes can directly access I/O servers through the network.*

PVFS is a parallel file system that provides high-speed access to file data for parallel applications. In addition, PVFS provides a clusterwide consistent name space, enables user-controlled striping of data across disks on different I/O nodes, and allows existing binaries to operate on PVFS files without the need for recompiling. File striping is illustrated in Figure 2.

Like many other parallel and cluster file systems, PVFS is designed as a client-server system with multiple servers, called I/O daemons. I/O daemons typically run on separate nodes in the cluster, called I/O nodes that have disks attached to them. PVFS is built on the local file system, which allows the Linux buffer cache to reduce the cost of individual local disk operations on the I/O servers. Each PVFS file is striped across the disks on the I/O nodes. Application processes interact with PVFS via a client library or by mounting the file system.

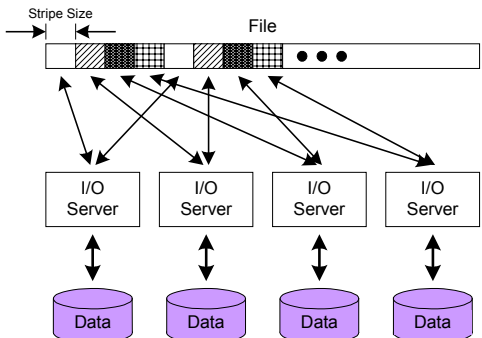

Figure 2. **Example file striping in PVFS**. *Files in PVFS can be striped according to user parameters that define the beginning I/O node, the number of I/O nodes to use, and the stripe size for data.*

PVFS also has a manager daemon that handles only metadata operations such as file permissions, file size, striping size, striped data location on disks, etc. When application processes (clients) open a PVFS file, the PVFS manager informs them of the locations of the I/O daemons directly. The manager does not participate in read/write operations; the client library and the I/O daemons handle all file I/O without the manager's intervention. This approach helps to minimize the impact of this potential bottleneck. The clients, I/O daemons, and the manager need not be run on different machines. However, running them on different machines will probably result in higher performance [6]. PVFS also supports MPI-I/O, the I/O chapter of the Message Passing Interface (MPI) 2 standard, through use of ROMIO [12].

By default, users must make multiple requests of PVFS in order to obtain noncontiguous data, one request per contiguous region. Thus, the number of contiguous I/O calls increases linearly with the number of contiguous regions in the noncontiguous request. We desire to create support in the file system for optimized noncontiguous access that significantly reduces the number of I/O requests.

## 3. Noncontiguous Data Access

Noncontiguous data access is an access that works on data that is not adjacent within file, memory, or both. The various types of noncontiguous data access are shown in Figure 3. One example of contiguous data in memory and noncontiguous data in file is an application that stores a two-dimensional array in a file, and then later desires to read the one element from each column into a contiguous memory buffer.

The more interesting of the noncontiguous data access patterns are the ones where the file data is noncontiguous. In order to optimize access when the file data is contiguous, a memory operation can buffer the access so that data movement is executed in memory and only one file read/write request is necessary. When the file is noncontiguous, buffering alone is not adequate. Other methods must be used to perform a noncontiguous data access when the file data is noncontiguous.

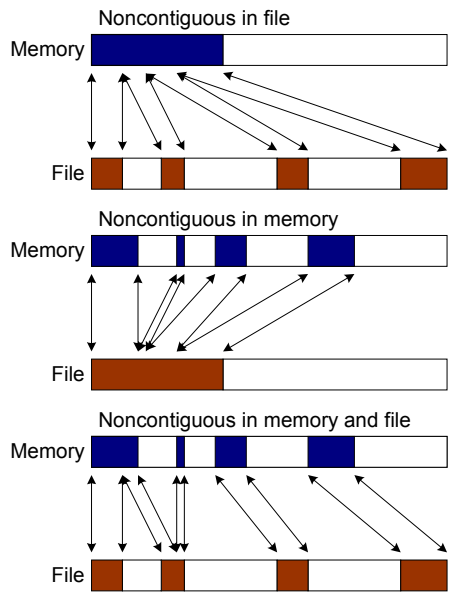

Figure 3. **Possible noncontiguous data accesses.** *This figure shows that noncontiguous data access can refer to noncontiguity in file, memory, or both.*

Often, studies have concluded that scientific applications access many small noncontiguous regions of data from a file [1][4][7][11][12]. If contiguous I/O requests must be used to perform these data accesses, large overhead resulting from multiple I/O requests will considerably hurt application run times. MPI-IO allows users to describe noncontiguous data access patterns but is limited in its ability to improve application performance if support for noncontiguous access is not present at the file system level.

Current solutions to the noncontiguous access pattern problem involve calling multiple independent I/O requests or using "data sieving" I/O techniques to take advantage of the high transfer rate of larger disk operations [13]. In this section we describe these two solutions for noncontiguous data access as well as our list I/O solution implemented in PVFS.

### 3.1 Multiple I/O

The interface to most parallel file systems allows for access only to a contiguous file region in a single I/O operation. Making multiple I/O operations performs the required noncontiguous access, as shown in Figure 4, but does so with a large cost of transmitting and processing I/O requests as well as many potential disk accesses (caching on the server side can alleviate this problem). We refer to this approach of handling noncontiguous access as *multiple I/O*. With thousands of compute nodes each making thousands of independent I/O operations, I/O servers must spend many of their CPU cycles processing new requests instead of delivering I/O to their clients. As clusters move to thousands or more processors, the I/O request problem grows worse.

### 3.2 Data Sieving I/O

Depending on the type of noncontiguous access pattern, "data sieving" may help to perform faster I/O in certain cases [13]. The data sieving approach handles noncontiguous access by moving a large region of data from file to a memory buffer, called the *data sieving buffer*, and performing the necessary data movement operations in memory at the client. An example use of data sieving is shown in Figure 5 where each I/O request covers several contiguous data regions. We chose to set the data sieving buffer at 32 MB for our testing purposes. For noncontiguous writes, using data sieving requires the file system to do a read-modify-write operation. A large section of data is read into the data sieving buffer, where the relevant regions are updated by write requests, and then the large contiguous section is written back to disk.

The advantage of using data sieving to perform noncontiguous data accesses is that multiple noncontiguous accesses can be described by a single I/O request. If the noncontiguous regions are nearby, the data sieving approach can eliminate many I/O requests. The data sieving approach can perform poorly, however, if the noncontiguous regions are far apart on disk. This access pattern will cause the single disk read to access a large amount of unused data that must move over the network. In general, using data sieving to perform noncontiguous I/O can benefit the user for noncontiguous access patterns that have relatively densely packed regions of desired data.

### 3.3 List I/O

PVFS has traditionally supported only contiguous requests for data. To address the performance problems inherent in the access patterns of scientific applications, we have added support for noncontiguous requests in PVFS. We desired a noncontiguous implementation that would reduce I/O accesses independent of the actual location in file. Based on the interface proposed [12], our implementation of noncontiguous data access, *list I/O*, would need support to describe any noncontiguous I/O pattern. The I/O servers would require support to process this request appropriately. The user would view the list I/O interface as follows:

```
pvfs_read_list(int mem_list_count,
               char *mem_offsets[ ],
               char mem_lengths[ ],
               int file_list_count,
               int file_offsets[ ],
               int file_lengths[ ])
```

(and similarly for pvfs_write_list).

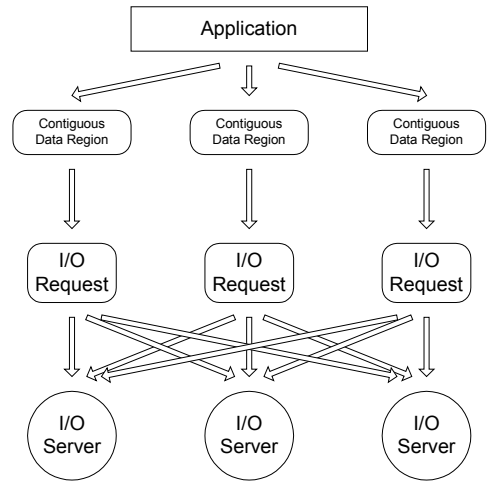

Figure 4. **Multiple I/O dataflow for noncontiguous I/O**. *In the multiple I/O approach, each noncontiguous data region requires a separate I/O request that the I/O servers must process.*

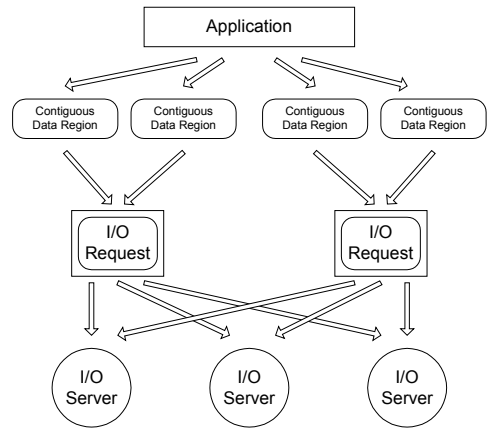

Figure 5. **Data sieving I/O dataflow for noncontiguous I/O**. *Noncontiguous data regions can sometimes be combined to reduce the number of I/O requests.*

*Mem_list_count* holds the total number of contiguous memory locations involved in the noncontiguous access. Similarly, *file_list_count* is the corresponding number of contiguous file locations. *Mem_offsets* is an array that references the beginning of each memory region, and the *mem_lengths* array matches each of these references with the corresponding memory lengths. *File_offsets* and *file_lengths* do the same for file regions.

PVFS clients make I/O requests through the PVFS library. These I/O requests contain information pertaining to a file (metadata, striping parameters) and can ask the I/O servers to perform operations such as read, write, open and close. In order for the I/O request to convey the description of noncontiguous data, we added another field to the I/O request structure to let the I/O servers know that a variable sized trailing data would follow the I/O request. This trailing data contains the file offsets and file lengths of the noncontiguous I/O request.

We modified the I/O server code to correctly process this routine by adding support to receive the trailing data and complete the I/O accesses. We have chosen to allow up to 64 contiguous file regions to be described in trailing data before another I/O request must be issued. Therefore, I/O requests that contain more file regions than the trailing data limit are broken up into several list I/O requests. This limit was chosen to allow the I/O request and trailing data to travel through the network in a single Ethernet packet (1500 bytes). This is a conservative limit that allows us to see how this approach might be used in a real system. Figure 6 illustrates the list I/O execution flow.

### 3.4 Analysis of Different Approaches

For multiple I/O and list I/O, disk accesses will vary if the memory regions align with the file regions. For data sieving I/O, the number of I/O requests is also dependent on the location of the physical data.

Multiple I/O has large disadvantages when compared with other noncontiguous access methods because of the large number of I/O requests generated by a single noncontiguous I/O operation. This approach leads to a huge overhead in transmitting and processing each request. The best access pattern for multiple I/O is one where there are only a few contiguous regions of data to be accessed in both memory and file.

Data sieving I/O can be a very efficient solution because of the low number of I/O requests for physically close noncontiguous data. The major disadvantage associated with data sieving I/O is the retrieval of useless accessed data that will have to flow over the network. Another slight drawback with data sieving is evident in the write cases, where read-modify-write must be used if the file region is noncontiguous. The ideal I/O pattern for showcasing data sieving I/O is one where there are many noncontiguous file regions and the gap between two successive regions is small.

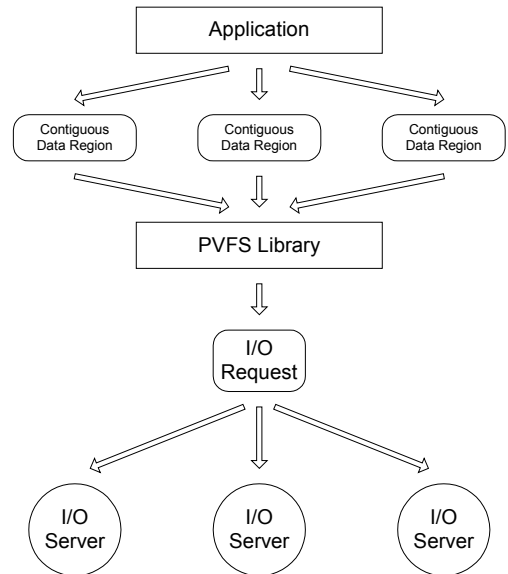

Figure 6. **List I/O dataflow for noncontiguous I/O**. *Noncontiguous data regions are described in a single I/O request.*

List I/O reduces the number of I/O requests in a noncontiguous data access by describing multiple file regions in a single list I/O request. Except for the case when noncontiguous regions are close enough for data sieving benefits to overcome the advantages of list I/O, list I/O will perform better than data sieving I/O.

## 4. Benchmark Overview and Results

This section will describe the tools used to gauge the performance of the list I/O optimization. We discuss the machine configuration, benchmarks, parameters that were chosen, and the experimental results.

### 4.1 Machine Configuration

We obtained all performance results on the Chiba City cluster at Argonne National Laboratory [3]. The cluster was configured as follows at the time of our experiments. There were 256 nodes, each with two 500-MHz Pentium III processors, 512 Mbytes of RAM, a 9 Gbyte Quantum Atlas IV SCSI disk, a 100 Mbits/sec Intel EtherExpress Pro fast-ethernet network card operating in full-duplex mode, and a 64-bit Myrinet card (Revision 3). We used only the fast Ethernet for our testing purposes. These nodes are currently using Red Hat 7.1. The kernel version 2.4.9 was compiled for SMP use. We used vari-ous compute nodes and 8 PVFS I/O nodes. One of the I/O nodes doubled as both a manager and an I/O server. For all of our tests we used the default stripe size of 16,384 bytes.

## 4.2 Benchmarks

Three types of benchmarks are presented in this paper: an artificial benchmark, a FLASH I/O astrophysics application, and a tiled visualization I/O application.

### 4.2.1 Artificial Benchmark

We created an artificial benchmark in order to test the noncontiguous performance of parallel reads and writes. We set the aggregate data access at 1 GByte in order to access a meaningful amount of data and also to have a baseline comparison. We also kept the I/O nodes constant at 8, with one doubling as both a manager and an I/O daemon. The benchmark varies the number of clients, the number of accesses, and the data access pattern. The data access patterns used in the benchmark are the one-dimensional cyclic and the two-dimensional block-block as shown in Figure 7 and Figure 8, respectively. Increasing the number of accesses further fragments the data access, making it more noncontiguous while preserving the aggregate data size. Changing the number of clients also determines the fragmentation of data. More clients accessing the same amount of data means more noncontiguity. The parallel reads and writes were conducted three times, and the I/O request time was averaged over the three runs. Because of the large execution time of multiple I/O in the write cases, however, we ran those tests only once. We decided not to use data sieving I/O with the parallel writes since data sieving requires a read-modify-write and therefore requires synchronization in which only one processor can write at a time in order to ensure the written data will not encounter any race conditions.

**One-Dimensional Cyclic** - This access pattern is a variable-grained, interleaved access, where we merge data from many processes into a single file in a cycling manner. An example of an application that would use this type of access pattern is one in which there is a global two-dimensional array and each processor operates on a region of columns of the array, as shown in Figure 7.

In these tests we vary the block size while maintaining a constant file size. Thus, a decrease in the block size increases the number of I/O requests for using multiple I/O. List I/O performance is expected to decrease as the accesses increase because of the need for additional requests, but not as rapidly as multiple I/O. Since the actual amount of data read is the same regardless of the number of accesses, we expect data sieving I/O to perform in a near constant time throughout the range of accesses. Note that as we increase the number of clients, data sieving I/O will be reading more and more useless data because the fraction of desired data in the accessed region decreases.

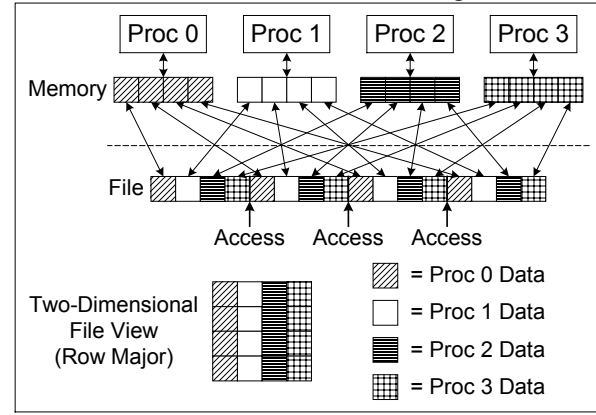

Figure 7. **Example one-dimensional cyclic access.** *An entire file stores a two-dimensional array and each processor is in charge of an equal amount of columns. The file view is also flattened into one-dimension.*

**Block-Block** - This type of access has a data distribution where a two-dimensional global array is partitioned by creating a block for every processor and organizing the blocks as shown in Figure 8. The tile application described later in section 4.4.1 uses an access pattern similar to this one.

### 4.2.2 Artificial Benchmark Results

For the one-dimensional cyclic access pattern, we expect linear results from both multiple I/O and list I/O since the number of I/O requests will increase linearly with the number of accesses. Data sieving should perform slightly better using the block-block access pattern due to the fact that the useful data is closer, which means accessing less impertinent data.

**One-Dimensional Cyclic Results** - Figure 9 shows that multiple I/O and list I/O scale linearly with the number of accesses. As we increase the number of accesses, the number of contiguous regions also increases, but the size of each contiguous region becomes smaller. Multiple I/O has to increase the number of I/O requests for a larger number of accesses. List I/O must also increase the number of I/O requests for a larger number of accesses, but at a slower rate than multiple I/O. Since list I/O can describe 64 file offsets and lengths in a single I/O request, list I/O will not be as affected as multiple I/O by a larger number of accesses.

We also notice that data sieving I/O stays fairly constant among any number of accesses for a fixed number of clients. This is because data sieving is moving the same amount of data in all of those cases. Also as expected, the time nearly doubles with data sieving I/O when the clients double due to the doubling of impertinent data read by each client (since each client now only has half as much relevant data in the same overall file region).

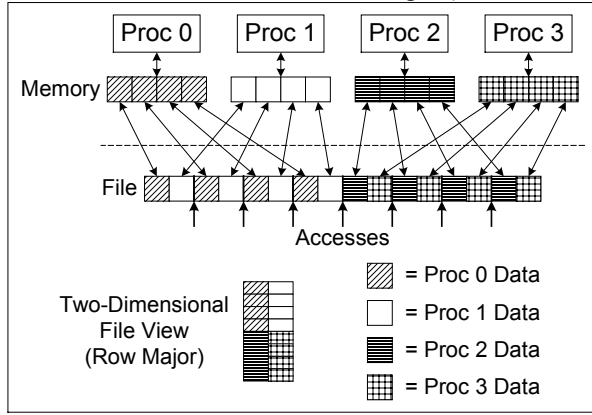

Figure 8. **Example block-block access**. *An entire file stores a two-dimensional array of blocks, and each processor is in charge of a single block. The file view has been flattened into 1-dimension.*

The write performance illustrated in Figure 10 for the one-dimensional cyclic access pattern is very poor for multiple I/O. Throughout most of the figures we can see that list I/O and multiple I/O have a performance gap of nearly two orders of magnitude. Both list I/O and multiple I/O performance degrades with the number of accesses but maintain their two order magnitude difference.

**Block-Block Results -** The results described in Figure 11 in the block-block read tests showed the trend expected for multiple I/O and data sieving I/O. Multiple I/O increases at a linear rate with the number of accesses while data sieving I/O remains nearly constant among the range of accesses.

List I/O performs unusually in the evaluation of 9 and 16 client block-block reads. When using 4 clients to read a file in a block-block distribution, list I/O scales up linearly with the number of accesses. However, we note that in Figure 11 for 9 and 16 clients, the list I/O curve sharply turns upward at some number of accesses. For 9 clients, each access is of size (1024*1024*1024 bytes)/(9 clients)/(800,000 accesses) $\approx$ 149 bytes/access at the turning point. Due to the block-block access pattern, each client heavily uses only a fraction of all the I/O servers, unlike the one-dimensional cyclic access pattern, which distributes a compute node's I/O load over all the I/O servers. Increasing the number of accesses for the block-block access pattern doesn't spread out the load as in the one-dimensional cyclic case. We observed the greater increase of list I/O with the number of accesses in the block-block access pattern at about 150 bytes/access for both the 9 client and 16 client cases.

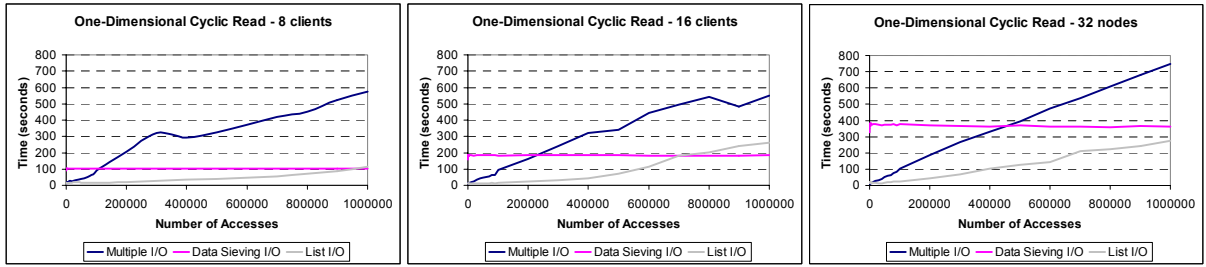

Figure 9: **One-dimensional cyclic read results with various clients**. *These results are obtained by using 8-32 clients reading data with the one-dimensional cyclic file access pattern.*

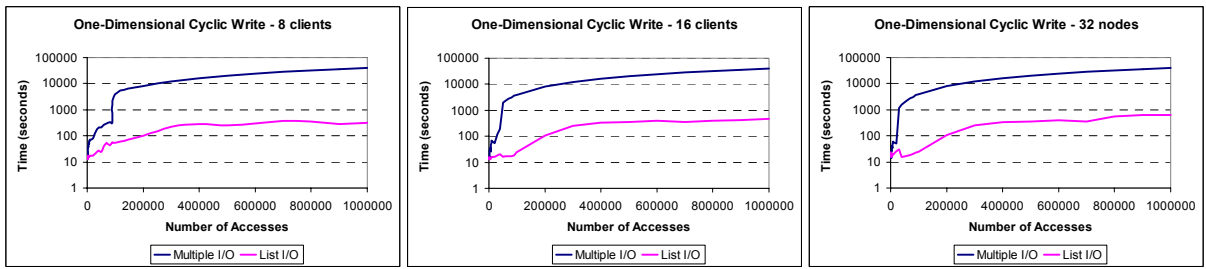

Figure 10: **One-dimensional cyclic write results with various clients**. *These results are obtained by using 8-32 clients writing data with the one-dimensional cyclic file access pattern.*

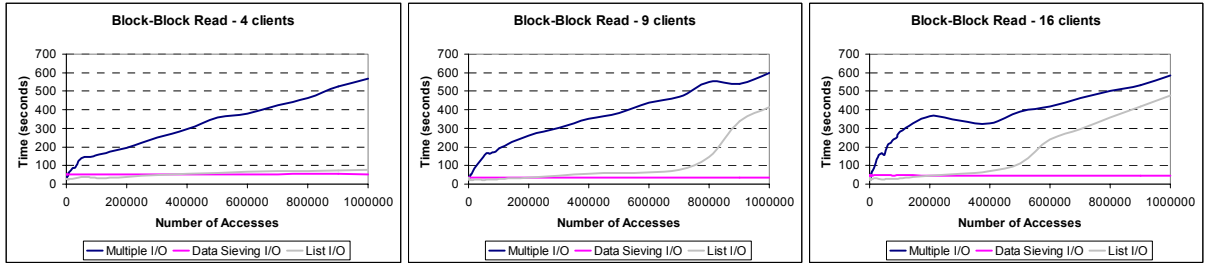

Figure 11. **Block-block read results with various clients**. *These results are obtained by using 4-16 clients reading data with the block-block file access pattern.*

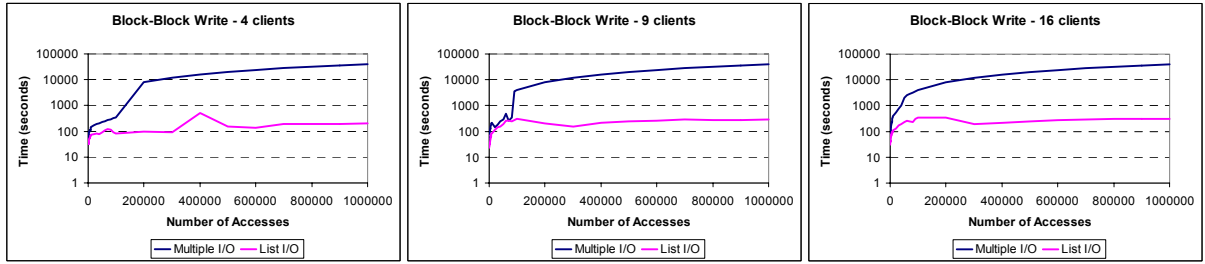

Figure 12. **Block-block write results with various clients**. *These results are obtained by using 4-16 clients reading data with the block-block file access pattern.*

A comparison between the 16 node cases in Figure 9 and Figure 11 show that the data sieving I/O times are reduced. The reason is that the data sieving I/O accesses less irrelevant data using the block-block access pattern.

Figure 12 shows that the block-block write results perform similar to the one-dimensional cyclic write results for multiple I/O and list I/O. As the number of accesses increases, multiple I/O and list I/O run times increase while maintaining the two orders of magnitude difference. The trend follows the results of the writes of the one-dimensional cyclic case.

### 4.3.1 FLASH I/O Benchmark

The FLASH code is an adaptive mesh refinement application that solves fully compressible, reactive hydrodynamic equations, developed mainly for the study of nuclear flashes on neutron stars and white dwarfs [5]. The I/O requirement for such an application often accounts for much of the running time. Instead of running the entire FLASH code, we simulate the I/O checkpoint writes where the element data in every block on every processor is written to file by using PVFS library calls. The access pattern of the FLASH code is noncontiguous both in memory and in file, making it a challenging application for parallel I/O systems. The FLASH memory structure consists of 80 FLASH three-dimensional blocks, or cells in the refined mesh, on each processor. Every block contains an inner data block surrounded by guard cells. Each of these data elements has 24 variables associated with it. Every processor writes these blocks to a file in a manner such that the file appears as the data for variable 0, then the data for variable 1, all the way up to variable 23. Within each variable, there exist 80 blocks, each of these blocks containing all the FLASH blocks from every processor [8]. The memory distribution is sketched in Figure 13, and the file hierarchy is explained in Figure 14.

The variable parameter in our implementation of the FLASH I/O benchmark is the number of clients. We varied the number of clients in the FLASH I/O benchmark while holding the other parameters consistent with the FLASH defined parameters. We ran each test once for 4-32 clients and used each noncontiguous method.

Each contiguous memory region is only the size of a double (8 bytes). In file, however, the contiguous regions are (8 x-elements)*(8 y-elements)*(8 z-elements)*(sizeofdouble) = 4096 bytes.

The FLASH I/O code is worst for multiple I/O. The number of I/O requests for multiple I/O = (80 blocks)*(8 x-elements)*(8 y-elements)*(8 z-elements)*(24 variables) = 983,040 I/O requests / processor. Our implementation of list I/O can do a little better, since list I/O describes the file regions, list I/O can reduce the amount of I/O requests to (80 blocks)*(24 variables)/64 = 30 I/O requests/processor, given the limit we placed on the number of regions described in a single request. Data sieving I/O can easily provide the fewest amount of requests, since the data size is only (80 blocks)*(8 x-elements)*(8 y-elements)*(8 z-elements)*(24 variables)*(sizeofdouble) = 7,864,320 bytes/processor (which is smaller than our data sieving buffer size of 32 MB), we need to make only one I/O request/processor. Data sieving writes, however, have the property that if multiple processors are writing to the same file and the file cannot be locked, there may be a chance of a race condition producing unpredictable results. Since there is no file locking mechanism in PVFS, we used an MPI_Barrier() to serialize access between processor data sieving writes. We implemented this through a for loop and a barrier so that during every iteration of the for loop, only one of the processors will write and then synchronize and then another processor will write and then synchronize until all the processors have written their data to file.

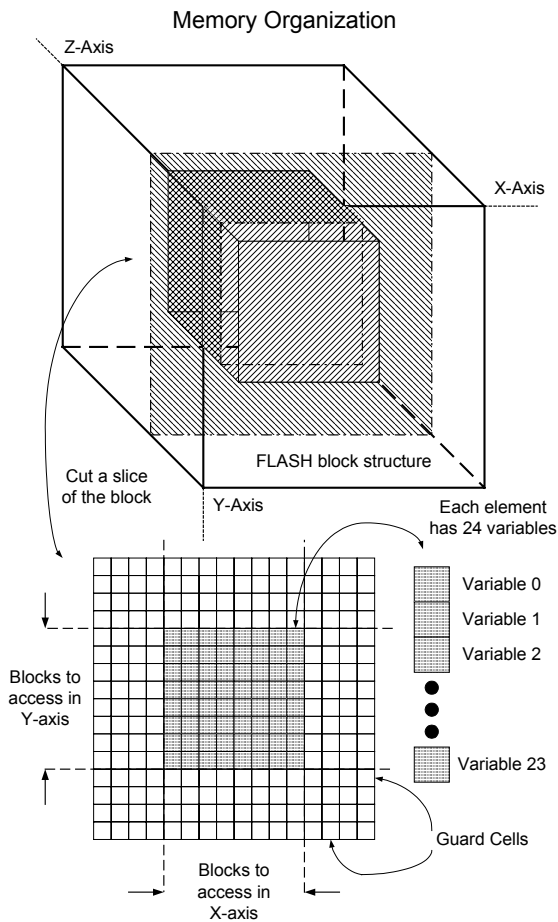

Figure 13. **FLASH memory organization**. *Each processor holds 80 FLASH blocks. Each of these blocks comprises a 8x8x8 cube of elements surrounded by guard cells. Within every element of data there are 24 double sized variables. As we scale up the number of processors, the total amount of data scales up as well.*

### 4.3.2 FLASH I/O Benchmark Results

In our FLASH I/O tests, the data size per compute process was fixed, while the number of compute processes was varied from 2 to 32. The file size increases linearly with the number of compute processes. Every additional compute node adds an additional 7.5 MBytes to the file. We held all other parameters consistent with the actual FLASH code. As described in Section 4.3.1, we expect data sieving to perform the best in this environment.

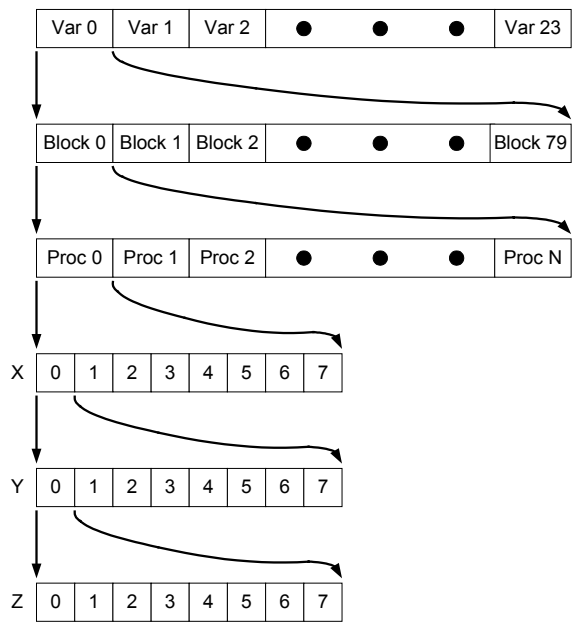

Figure 14. **FLASH file organization**. *The FLASH file organization is much different from the memory organization. At the highest level of viewing the file, all of the variable 0's are contiguous, then the variable 1's, etc. Within the chunk of a variable, there are 80 FLASH blocks. Every block contains the pertinent FLASH block from every processor.*

The FLASH I/O test highlights the power of the data sieving optimization. By combining I/O requests through buffering, we expect performance significantly better than list I/O and many orders of magnitude better than multiple I/O, both of which fail under the sheer number of noncontiguous regions. Because of the read-modify-write access method of data sieving I/O writes, file access must be serialized across compute processes. Again, we implemented these semantics through MPI_Barrier(), which should slow data sieving I/O performance slightly.

Multiple I/O and list I/O performed fairly consistently regardless of the number of clients, since PVFS scales without problems for these numbers of clients. Data sieving I/O time increases with the number of clients because of the file synchronization and a growing amount of useless data being accessed. Since a larger number of clients means more separation of useful data, data sieving I/O will not be as effective as when the relevant data was closer together.

Figure 15 shows the considerable impact of large amounts of I/O requests in the multiple I/O results. List I/O is approximately two orders of magnitude slower than data sieving I/O and a little over one order of magnitude faster than multiple I/O. The FLASH I/O results show that data sieving I/O can be very useful for this type of access pattern. Since multiple I/O requires so many small I/O requests, it cannot compete with data sieving I/O. List I/O also has the problem of a large number of small I/O accesses, but to a lesser degree.

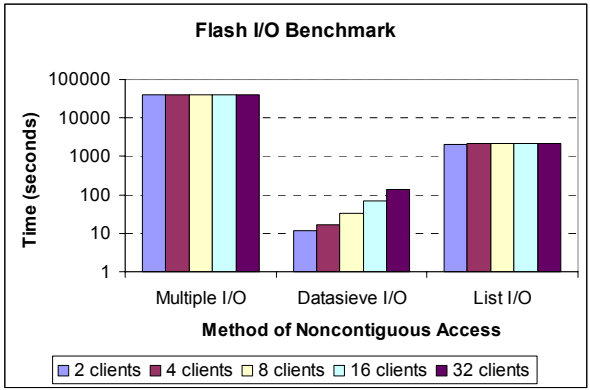

Figure 15. **FLASH I/O results**. *This test shows that even with the file locking slowdown, data sieving I/O can easily outperform the other two noncontiguous methods.*

**4.4.1 Tiled Visualization I/O**

The tiled visualization code takes display files and divides them into blocks of files, creating an array of displays. This type of display division is useful for viewing high-resolution playback on a larger screen by using more displays. In our benchmark, we implement the I/O portion of the code in terms of PVFS library calls, where multiple compute nodes read a single file and each node accesses a part of that data in the pattern shown in Figure 16. We monitor the time to take a large display file and read the relevant tile data to each respective node.

For our tests, we used 6 compute nodes along with the 8 node PVFS setup. The parameters were a 3 x 2 display with each display rendering 1024 x 768 with 24 bit color. Between the displays there was a 270 pixel horizontal overlap and a 128 pixel vertical overlap, bringing the file size to about 10.2 MBytes. Each test was run three times and the data was averaged.

In the tiled visualization I/O code, we expect that the list I/O code will outperform the other noncontiguous methods. Since file access is noncontiguous (but each chunk is fairly large) and the memory is contiguous, list I/O will need to perform a minimal number of I/O re-quests (768/64 = 12). Multiple I/O requires 768 I/O requests because of its description of only a contiguous region. Data sieving I/O should perform reasonably, but in this case the client will end up using only a fraction (1 / number of tiles in the x direction, for this case 1/3) of the actual data read.

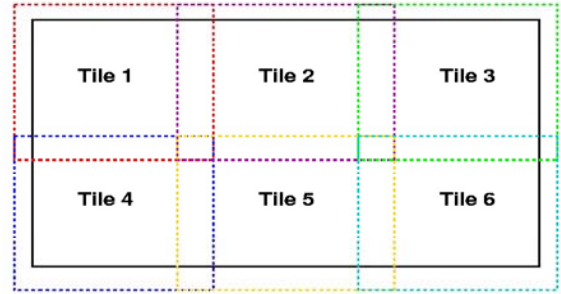

Figure 16. **Tiled visualization I/O access pattern**. *Each processor holds a "tile" of the actual file. The file is organized in a row major order. For this example, the beginning of the file would be proc 0 data, proc 1 data, proc2 data, then proc 0 data, proc 1 data, proc 2 data, etc.*

**4.4.2 Tiled Visualization I/O Benchmark Results**

The tiled visualization access pattern is actually very similar to the one tested in the artificial benchmark in the block-block case. As we expected, we see that list I/O is able to perform more than twice as well as either of the other two methods.

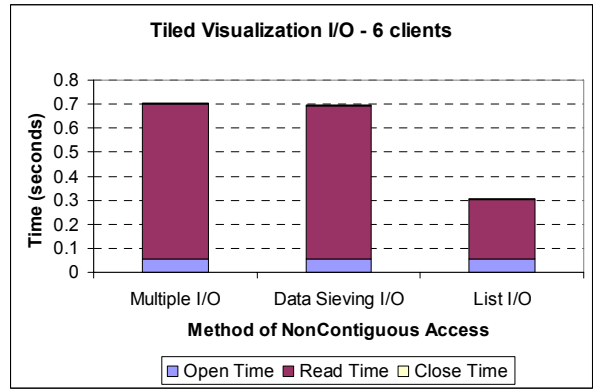

Figure 17. **Results of the tiled visualization I/O benchmark with 6 clients**. *In this benchmark, we indicate the time to open the file, have each processor perform the read, and the close file time.*

## 5. Conclusion

Our experimental results have shown that list I/O can perform noncontiguous data access much faster than tradi-tional methods for some significant workloads. In situations where most of the noncontiguous regions are close together, data sieving produces better results. Overall, multiple I/O should not be considered for large-scale scientific applications with noncontiguous accesses patterns.

A combination of the list I/O and data sieving approaches could provide a hybrid solution that would be applicable over a larger range of access patterns. More research should be conducted on ways to use data sieving techniques in conjunction with the list I/O implementation. For example, if two noncontiguous regions are close to each other, a data sieving operation may take place for just those particular regions. Determining such a case may, however, require more complex software design and suffer considerable overhead.

Even more interesting is the possibility of using more descriptive languages for identifying requested regions. In all of our tests, the access patterns are regular. Support for I/O requests that use an approach similar to MPI datatypes, for example, would describe these patterns with vector datatypes. This would eliminate the linear relationship between the number of contiguous regions and the number of I/O requests. By doing so, the largest drawback of the list I/O approach could be avoided.

## Acknowledgments

This project was sponsored by the Scientific Data Management Center of the DOE SCIDAC program, the DOE ASCI program and a grant from the National Science Foundation. We also enjoyed the support of the Mathematical, Information and Computational Sciences Division subprogram of the Office of Advanced Scientific Computing Research, U.S. Department of Energy, under Contract W-31-109-Eng-38.